\documentclass[twocolumn,floatfix,prb,aps,showpacs,longbibliography]{revtex4-2}
\usepackage{graphicx,amsmath,amssymb,color}
\usepackage{nicefrac}
\usepackage[titletoc,title]{appendix}
\usepackage[colorlinks,bookmarks=true,citecolor=blue,linkcolor=red,urlcolor=blue]{hyperref}

\begin{document}

\title{Parton wave function for the fractional quantum Hall effect at $\nu=6/17$}
\author{Ajit C. Balram$^{1}$ and A. W\'ojs$^{2}$}
\affiliation{$^{1}$Institute of Mathematical Sciences, HBNI, CIT Campus, Chennai 600113, India}
\affiliation{$^{2}$Department of Theoretical Physics, Wroc\l{}aw University of Science and Technology, 50-370 Wroc\l{}aw, Poland}
\date{\today}

\begin{abstract} 
We consider the fractional quantum Hall effect at the filling $\nu=6/17$, where experiments have observed features of incompressibility in the form of a minimum in the longitudinal resistance. We propose a parton state denoted as ``$3\bar{2}1^{3}$" and show it to be a feasible candidate to capture the ground state at $\nu=6/17$. We work out the low-energy effective theory of the $3\bar{2}1^{3}$ edge and make several predictions for experimentally measurable properties of the state which can help detect its underlying topological order. Intriguingly, we find that the $3\bar{2}1^{3}$ state likely lies in the same universality class as the state obtained from composite-fermionizing the 1+1/5 Laughlin state. 
\end{abstract}

\maketitle

\section{Introduction}
Two-dimensional electrons subjected to a strong perpendicular magnetic field and cooled to ultra-low temperatures arrange themselves in some of the most exotic strongly correlated states realized in nature. This marvelous collective phenomenon, known as the fractional quantum Hall effect (FQHE)~\cite{Tsui82}, is a paradigm for studying interacting electronic systems that cannot be connected by perturbation theory to free electrons. In the lowest Landau level (LLL) FQHE primarily occurs at filling factors $\nu=n/(2pn\pm 1)$, where $n$ and $p$ are positive integers. These states can be explained as a manifestation of the integer quantum Hall effect (IQHE) of composite fermions (CFs)~\cite{Jain89} which are topological bound states of electrons and an even number ($2p$) of quantized vortices. A major portion of the FQHE phenomenology that occurs in the LLL can be well-understood using the theory of non-interacting CFs~\cite{Jain07}. Nevertheless, in the range of fillings $1/3<\nu<2/5$ very high-quality samples exhibit signatures of FQHE at $\nu=4/11,~5/13,~6/17,~3/8,~3/10,~4/13$, and $5/17$~\cite{Pan03, Pan15, Samkharadze15b,Kumar18a,Chung21}. An understanding of these fragile states necessitates going beyond the framework of weakly interacting composite fermions as these states likely arise from an FQHE of CFs themselves~\cite{Mukherjee12, Mukherjee14c, Mukherjee15b, Balram16c}.  This article focuses on one such state, namely the $6/17$ FQHE.

Amongst the aforementioned delicate states, incompressibility has already been established at $4/11$ and $5/13$. Experiments by Pan \emph{et al.}~\cite{Pan15} provided the first evidence for the existence of a gapped state at $4/11$. Subsequently, Samkharadze \emph{et al.}~\cite{Samkharadze15b} also confirmed the formation of an incompressible state at $4/11$.  Samkharadze \emph{et al.}~\cite{Samkharadze15b} further showed that incompressibility also occurs at $5/13$. However, they found that at the lowest accessible temperatures 6/17 remains compressible.  Encouragingly, recent experiments on ultra-high-mobility samples show strong features of FQHE in the form of deep longitudinal resistance minimum at many fractions between 1/3 and 2/5 including possibly at 6/17~\cite{Chung21}. These results suggest that just like at 4/11 and 5/13 the ground state at 6/17 could be an FQHE state. It would be interesting to study in detail the nature of the 6/17 state in these ultra-pure samples in the future. We note that although the 6/17 state may be non fully spin-polarized~\cite{Balram16c}, in this article, we focus only on the nature of a fully polarized state that could occur at this filling.

In the CF theory, the $\nu=6/17$ state of electrons is described by the $\nu^{*}=6/5=1+1/5$ state of CFs [the electronic filling $\nu$ is related to the CF filling $\nu^{*}$ as $\nu=\nu^{*}/(2\nu^{*}+1)$]. The $6/5$ state of CFs, in turn, arises from filling the lowest CF-Landau-like level [called the Lambda level ($\Lambda$L)] and creating a 1/5 FQHE state of CFs in the second $\Lambda$L (S$\Lambda$L). The natural expectation that CFs form a 1/5 Laughlin state~\cite{Laughlin83} in the S$\Lambda$L is not readily supported by numerical calculations~\cite{Mandal02, Mukherjee14}. The pseudopotentials $\{V_{m}\}$ ($V_{m}$ denotes the energy penalty for placing two particles in the relative angular momentum $m$ state) of CFs~\cite{Sitko96, Wojs00, Lee01, Lee02, Wojs04, Balram16c, Balram17b} indicate that the residual interaction between two CFs occupying the S$\Lambda$L in the presence of the filled lowest $\Lambda$L (L$\Lambda$L) is hollow-core like, i.e., the interaction is dominated by a strong repulsion in the $m=3$ channel.  For comparison, the Coulomb interaction between electrons in the LLL, which stabilizes the 1/5 Laughlin state~\cite{Ambrumenil88, Kusmierz18, Balram20a}, is dominated by the hard-core $V_{1}$ pseudopotential.  Thus, for interactions where $V_{3}\gg V_{m}~\forall m\neq 3$, the case realized for CFs in the S$\Lambda$L, it is not clear if the 1/5 Laughlin state could be stabilized.  

Motivated by the form of the CF-CF interaction, W\'ojs, Yi, and Quinn (WYQ) proposed using the $V_{3}$-only interaction as a model Hamiltonian~\cite{Wojs04} to study the FQHE states arising from interacting CFs residing in their S$\Lambda$L.  Numerical studies of the WYQ model show that it produces an exact zero-energy ground state at filling factor $1/5$~\cite{Wojs04, Wojs09a}. This state,  which we shall refer to as the 1/5 WYQ state, is topologically different from the 1/5 Laughlin state since it occurs at a different shift~\cite{Wen92} in the spherical geometry. Note that the 1/5 Laughlin state is also an exact zero-energy state of the WYQ Hamiltonian but it occurs at a lower density than the 1/5 WYQ state.  Nonetheless, the gaps of the 1/5 WYQ state for the $V_{3}$-only Hamiltonian do not extrapolate to a positive value in the thermodynamic limit indicating that it is a compressible state~\cite{Jolicoeur17a}. Thus, even the WYQ model is inadequate to understand the $6/5$ state of CFs, and thereby an underlying mechanism for the $6/17$ FQHE is lacking. We note that these results suggest that the consideration of only the interaction between two CFs may be insufficient to determine the nature of the delicate FQHE states arising from a partial filling of the S$\Lambda$L. In principle, one can incorporate interaction between three-CFs and further higher-body correlations~\cite{Balram17b} but doing so in practice is an extremely challenging task.

In this work, instead of taking recourse to the CF theory, we take a different approach and directly propose a wave function for the $6/17$ state of electrons.  A recent work~\cite{Balram21} showed that a parton state~\cite{Jain89b} can be used to understand the $4/11$ FQHE. Building on that work, we construct a parton wave function at 6/17 and show that it provides a plausible representation of the numerically obtained Coulomb ground state in the LLL.  Furthermore, we use the low-energy theory of the $6/17$ parton edge to make predictions that could be tested out in future experiments.  Remarkably, we find that the topological properties of our parton state are identical to those of the $\nu^{*}= 1+1/5$ Laughlin state mentioned above which suggests that they are likely different microscopic manifestations of the same underlying phase.  At present we do not know of a way to directly interpret our parton state as a composite-fermionized (CFized) 6/5 state, where the CFs form a $1/5$ Laughlin liquid in the S$\Lambda$L.

The article is organized as follows. In the next section, Sec.~\ref{sec: parton_ansatz_6_17}, we provide a primer on parton states and introduce our ansatz for the 6/17 FQHE. Then in Sec.~\ref{sec: ED_comparison}, we present extensive results comparing our wave function with the LLL Coulomb ground state at 6/17 obtained from brute-force diagonalization. In Sec.~\ref{sec:eff_edge} we discuss topological properties of our state that are determined from a low-energy effective theory of its edge. We wrap up the paper in Sec.~\ref{sec: discussions} with a discussion of the experimental signatures that could reveal the parton order.

\section{Primer on parton states and the parton ansatz for 6/17}
\label{sec: parton_ansatz_6_17}
Generalizing the idea of composite fermions, Jain introduced the parton theory~\cite{Jain89b} to construct a larger class of FQHE wave functions. In the parton theory,  one imagines dividing the electron into $q$ fictitious objects called partons. To construct an incompressible state, each of the partons is placed in an IQHE state at filling factor $n_\gamma$, where $\gamma=1,2,\cdots, q$ labels the different parton species. The resulting state denoted as ``$n_1n_2n_3...$," is described by the wave function
\begin{equation}
\Psi^{n_1n_2n_3...}_\nu = \mathcal{P}_{\rm LLL} \prod_{\gamma=1}^{q}\Phi_{n_\gamma}(\{z_k\}),
\label{eq:parton_wf}
\end{equation}
where $z_{k}=x_{k}-iy_{k}$ is the complex representation of the two-dimensional coordinate of the $k$th electron, $\Phi_n$ is the wave function for the state with $n$ filled LLs of non-interacting electrons, and $\mathcal{P}_{\rm LLL}$ denotes projection into the LLL as is appropriate for the high field limit. The effective magnetic field seen by the partons can be anti-parallel to that experienced by the electrons. Such states correspond to negative fillings for the partons, which we denote by $\bar{n}$, with $\Phi_{\bar{n}}=\Phi_{-n}=\Phi_n^*$.  

The partons are unphysical entities so they should be stuck back together to recover the physical electrons. At the level of wave functions, this gluing procedure is implemented by setting the different parton species coordinates $z_k^\gamma$ equal to the electron coordinate $z_k$, i.e., $z_k^\gamma = z_k$ for all $\gamma$.  Thus, each Slater determinant wave function $\Phi_{n_\gamma}$ in Eq.~(\ref{eq:parton_wf}) is composed of \emph{all} the electronic coordinates $\{z_{k}\}$. As the density of each parton species is identical to the electronic density and all the partons are exposed to the same external magnetic field, the charge of the $\gamma$ parton species $e_\gamma$ is related to the electron's charge $-e$ as $e_\gamma = -\nu e / n_\gamma$. The constraint that the sum of the parton charges add up to the electronic charge implies that the electronic filling factor is given as $\nu = [\sum_{\gamma=1}^{q} n_\gamma^{-1}]^{-1}$ in terms of the parton fillings $\{n_{\gamma}\}$.  

Many notable classes of FQHE states can be obtained as special cases of parton states. The Laughlin state~\cite{Laughlin83} at $\nu=1/(2p+1)$, described by the wave function $\Psi_{\nu=1/(2p+1)}^{\rm Laughlin}=\Phi^{2p+1}_{1}$, can be viewed as the $(2p+1)$-parton state where each parton forms a $\nu=1$ IQHE state. The Jain states, described by the wave function $\Psi_{\nu=n/(2pn\pm 1)}^{\rm Jain}=\mathcal{P}_{\rm LLL}\Phi_{\pm n}\Phi^{2p}_{1}$, can be re-interpreted as $(2p+1)$-parton states where $2p$ partons form a $\nu=1$ IQHE state and one parton forms a $\nu=\pm n$ IQHE state.  Numerous parton states, beyond the Laughlin and Jain states, have been proposed as promising candidates to describe FQHE states occurring in the S$\Lambda$L~\cite{Balram21}, second LL (SLL)~\cite{Balram18, Balram18a, Balram19, Balram20, Balram20a, Balram20b, Faugno20b}, wide quantum wells~\cite{Faugno19},  and in the LLs of graphene~\cite{Wu17, Bandyopadhyay18, Kim19, Faugno20a, Balram20, Faugno20b}.  These works suggest that it is plausible that viable candidate parton states can be constructed to capture all the observed FQHE states~\cite{Balram19, Balram20a}.

For the 6/17 FQHE, we propose the parton state denoted as ``$3\bar{2}1^{3}$" and described by the wave function
\begin{equation}
\Psi^{3\bar{2}1^{3}}_{6/17} = \mathcal{P}_{\rm LLL} \Phi_{3}[\Phi_{2}]^{*}\Phi^{3}_{1} \sim \frac{\Psi^{\rm Jain}_{3/7}\Psi^{\rm Jain}_{2/3}}{\Phi_{1}}.
\label{eq: parton_6_17_3bar2111}
\end{equation}
This state is the $n=3$ member of the $n\bar{2}1^{3}$ sequence, the $n=4$ member of which was recently shown to be relevant for the $4/11$ FQHE~\cite{Balram21}. The $\sim$ sign in Eq.~(\ref{eq: parton_6_17_3bar2111}) denotes that the states on either side of the sign differ in details of how the projection to the LLL is implemented. We expect that such details do not change the universality class of the wave function~\cite{Balram16b}. The form of the $3\bar{2}1^{3}$ wave function given on the right-most side of Eq.~(\ref{eq: parton_6_17_3bar2111}) is the one that is amenable to a numerical evaluation and thus it is this form that we shall use in the numerical computations shown below.  On the spherical geometry, this state occurs at a shift~\cite{Wen92} of $\mathcal{S}=4$.  Moreover, it is the only known shift where the LLL Coulomb ground-state at 6/17 is consistently uniform and incompressible for all accessible finite systems~\cite{Balram16c}. 

Interestingly, the shift of the above parton state is the same as that of the 6/17 state which results from filling the L$\Lambda$L and forming the 1/5 Laughlin state in the S$\Lambda$L~\cite{Balram16c}.  The wave function for this state, which we called ``CFized $6/5$,'' is given by
\begin{equation}
\Psi^{{\rm CFized~6/5}}_{6/17} = \mathcal{P}_{\rm LLL} \Phi^{2}_{1}\Phi_{1+1/5},
\label{eq: CFized_6_5}
\end{equation}
where $\Phi_{1+1/5}$ denotes the state of electrons in which the LLL is filled and a $1/5$ Laughlin state is formed in the SLL.  We note that in comparison to the wave function given in Eq.~(\ref{eq: parton_6_17_3bar2111}), it is a daunting task to numerically construct the wave function given in Eq.~(\ref{eq: CFized_6_5}).  Nevertheless, we shall show below that these two states likely lie in the same topological phase. In contrast, the 6/17 state resulting from filling the L$\Lambda$L and forming the 1/5 WYQ state in the S$\Lambda$L occurs at a shift $\mathcal{S}=14/3$~\cite{Balram16c} and thus is topologically distinct from the $3\bar{2}1^{3}$ state.  The LLL Coulomb ground state for $N=14$ electrons is not uniform at this shift of $\mathcal{S}=14/3$~\citep{Balram16c} which suggests that the $6/17$ FQHE is unlikely to lend itself to a description in terms of a CFized $1+1/5$ WYQ state. 

\section{Numerical results}
\label{sec: ED_comparison}
For all our numerical computations we shall use Haldane's spherical geometry~\cite{Haldane83}. In this geometry, $N$ electrons move on the surface of a sphere in the presence of a radial magnetic field that produces a flux of strength $2Q(hc/e)$ ($2Q$ is a positive integer). The total orbital angular momentum $L$ and it's $z$-component $L_{z}$ are good quantum numbers in this geometry. Quantum Hall states on the sphere have $L=0$ and occur when $2Q=\nu^{-1}N-\mathcal{S}$, where $\mathcal{S}$ is a rational number called the shift~\cite{Wen92}. In particular, the shift~\cite{Wen92} of the parton states of Eq.~(\ref{eq:parton_wf}) is $\mathcal{S}=\sum_{\gamma=1}^{q} n_\gamma$. The IQHE state with $n$ filled LLs can only be constructed on the sphere when $N\geq n^{2}$ with $N$ being divisible by $n$. Thus, the $3\bar{2}1^{3}$ state of our interest can only be constructed on the sphere for $N=6l$, where $l$ is an integer with $l\geq 2$.  Only the smallest system with $N=12$ and $2Q=30$, which has a Hilbert space dimension of about 2.27 million, is accessible to exact diagonalization (ED). Below we shall present in detail our results obtained from ED on this system. The next system for which the $3\bar{2}1^{3}$ wave function can be constructed is that of $N=18$ electrons at flux $2Q=47$. This system has a dimension of over 61 billion and is well beyond the reach of ED. 

To compare the $3\bar{2}1^{3}$ state against results obtained from ED it is useful to have the Fock-space representation of the parton state. In particular, this facilitates a calculation of overlaps and energies of the trial state for a wide range of interactions. To obtain the second-quantized representation of the $3\bar{2}1^{3}$ wave function we use the method outlined in Refs.~\cite{Sreejith11, Balram18, Balram20a}.  Since the $3\bar{2}1^{3}$ state is uniform we first evaluate all the total orbital angular momentum $L=0$ states for the relevant system and then find the expansion coefficients of the desired wave function on this basis of all the $L=0$ states.  

To obtain an orthonormal set of basis states for the $L=0$ subspace we first evaluate dimension($L=0$) number of linearly independent $L=0$ states by starting with random initial vectors in the $L_{z}=0$ subspace and Lanczos diagonalizing the $L^{2}$ operator using these vectors.  Next, to orthogonalize the vectors obtained in the previous step we use the Gram-Schmidt orthogonalization procedure. Finally, if the orthogonalization spoils the $L=0$ value of any state, we rerun a few $L^2$ Lanczos iterations on it to ensure that the vector, in the end, has $L=0$. Correcting the vectors this way, we end up with a complete set of orthogonal $L=0$ vectors.

To find the expansion coefficients, we evaluate the wave function given in Eq.~(\ref{eq: parton_6_17_3bar2111}) as well as all the $L=0$ states for a sufficiently large number of configurations $\{z_{k}\}$ of the particles. This results in a set of linear equations the solution of which gives us the requisite expansion coefficients.  To numerically calculate the wave function given in Eq.~(\ref{eq: parton_6_17_3bar2111}) we obtain the constituent 12-particle 2/3 and 3/7 Jain states~\cite{Yang19a} by a direct brute-force projection of the respective unprojected Jain states to the LLL. 

The exact LLL ground state of the system of $N=12$ electrons at flux $2Q=30$ is uniform and has per-particle Coulomb energy, including the contribution of the positively charged background, of $-0.4361~e^{2}/(\epsilon\ell)$, where $\ell=\sqrt{\hbar c/(eB)}$ is the magnetic length, $B$ is the magnetic field and $\epsilon$ is the dielectric constant of the background host material. In comparison,  the Coulomb energy of the $3\bar{2}1^{3}$ state is $-0.4276~ e^{2}/(\epsilon\ell)$ which is within $2\%$ of the exact energy.  The overlap of the exact Coulomb ground state with the $3\bar{2}1^{3}$ wave function is 0.72.  This agreement, although not as exquisite as that between the exact LLL Coulomb ground state and the Laughlin or Jain states~\cite{Jain07, Balram13, Kusmierz18, Balram20b}, is on par with that of other candidate states that arise in the S$\Lambda$L~\cite{Balram21} and the SLL~\cite{Balram20b}.  

\begin{figure}[htpb]
\begin{center}
\includegraphics[width=0.47\textwidth,height=0.23\textwidth]{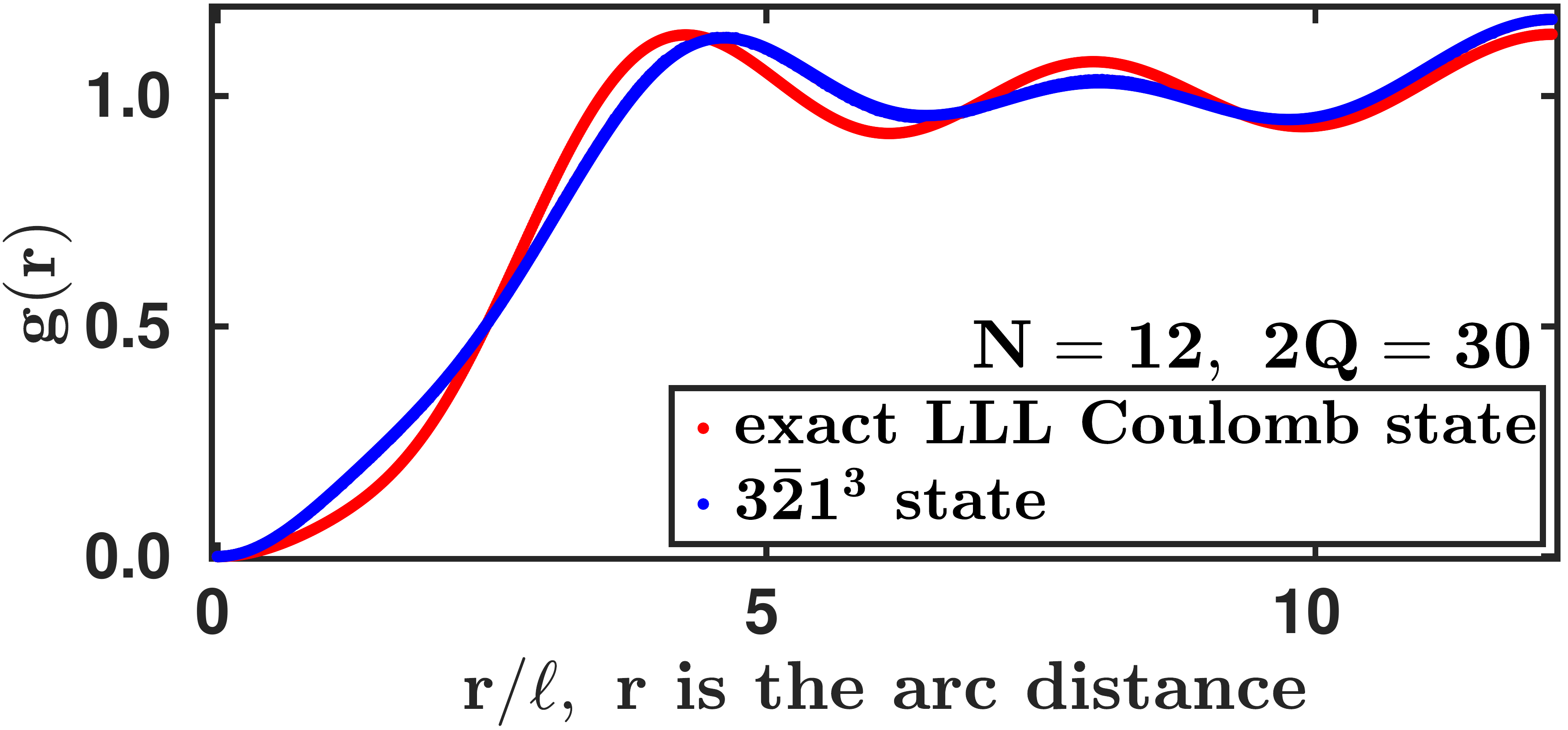} 
\caption{(color online) The pair correlation function $g(r)$ as a function of the distance $r$, measured in units of the magnetic length $\ell$, along the arc of the sphere for the exact lowest Landau level Coulomb ground state (red), and the $3\bar{2}1^{3}$ state of Eq.~(\ref{eq: parton_6_17_3bar2111}) (blue) for $N=12$ electrons at flux $2Q=30$.}
\label{fig: pair_correlations_6_17}
\end{center}
\end{figure}

In Fig.~\ref{fig: pair_correlations_6_17} we show a comparison of the pair-correlation function $g(r)$ of the exact Coulomb ground state and the $3\bar{2}1^{3}$ states. The $g(r)$ of the exact and $3\bar{2}1^{3}$ states are in good agreement with each other. Both states show decaying oscillations at long distances which is a typical fingerprint of gapped states~\cite{Kamilla97, Balram15b, Balram17}. While the $g(r)$ of the two states differ in the short-to-intermediate distance regime,  at long distances, the $g(r)$ of the $3\bar{2}1^{3}$ state nicely tracks the $g(r)$ of the exact state. Thus, although the $3\bar{2}1^{3}$ state does not provide an extremely accurate microscopic description of the LLL Coulomb ground state, it likely lies in the same topological phase as the true ground state. The $g(r)$ of the $3\bar{2}1^{3}$ and the Coulomb ground state lack a ``shoulder-like" feature at intermediate distances (seen in non-Abelian states~\cite{Read99, Hutasoit16, Balram20}) which suggests that these states are Abelian. 

\begin{figure}[htpb]
\begin{center}
\includegraphics[width=0.47\textwidth,height=0.23\textwidth]{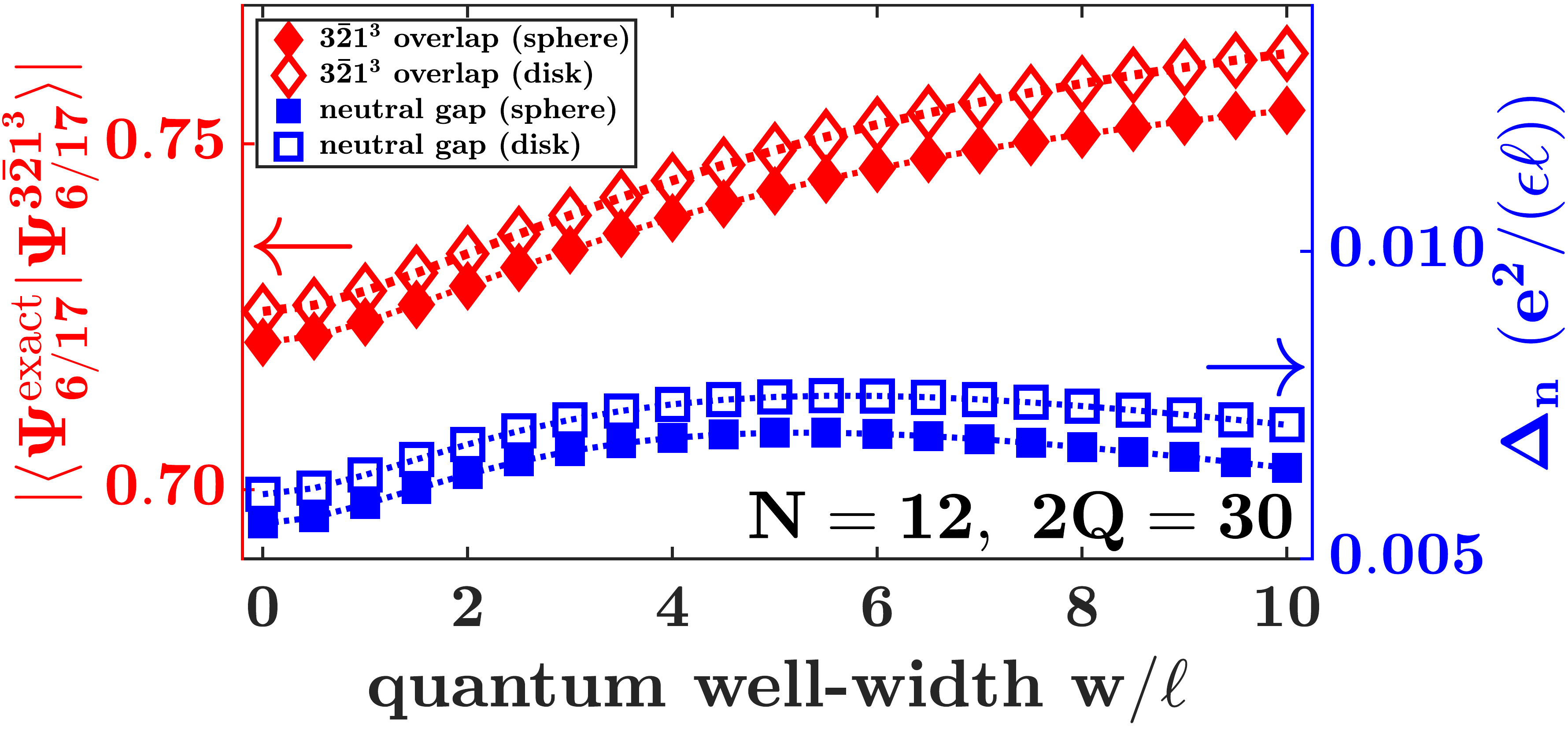} 
\caption{(color online) Overlaps of the $3\bar{2}1^{3}$ state with the exact lowest Landau level Coulomb ground state (red diamonds) and neutral gaps (blue squares) as a function of the well-width $w/\ell$.  Results are shown for the spherical (filled symbols) and disk (open symbols) pseudopotentials for a system of $N=12$ electrons at flux $2Q=30$.}
\label{fig: gaps_overlaps_6_17}
\end{center}
\end{figure}

To further assess the stability of the $3\bar{2}1^{3}$ state we have calculated its overlap with the Coulomb ground state incorporating finite well-width corrections. We take a simplified model to simulate the effect of finite-thickness in which the transverse wave function of the quantum well is taken to be that of the ground state of an infinite square well of width $w$.  To allow for further variations in the interaction, we have evaluated the exact ground state using both the spherical and planar disk pseudopotentials.  The overlaps of the $3\bar{2}1^{3}$ state with the exact ground state of this model finite-width interaction are shown in Fig.~\ref{fig: gaps_overlaps_6_17}. The overlaps are of the order 0.75 and increase as the well width is increased indicating that the finite thickness of the quantum well enhances the stability of the $3\bar{2}1^{3}$ state. We mention here that, taking into account the effect of finite-width, WYQ argued that the $6/17$ FQHE could arise from the formation of a 1/5 Laughlin state of CFs in the S$\Lambda$L~\cite{Wojs04}.  Our finite-width results lend further support to their assertion.

Next, we turn to evaluate gaps for the $6/17$ FQHE. Since only a single system is accessible to ED we have not been able to estimate the thermodynamic gaps of the $6/17$ state. Nevertheless, for the system of $N=12$ electrons, we have evaluated the neutral gap which is defined as the difference in the two lowest-energies at the flux of $2Q=30$. The neutral gaps, also shown in Fig.~\ref{fig: gaps_overlaps_6_17}, are of the order $0.005~ e^{2}/(\epsilon\ell)$ for the entire range of widths considered. Therefore, the ground state is robust to changes in the interaction stemming from the finite width of the quantum well.  For comparison, the neutral gap at the neighboring 1/3 state is an order of magnitude larger~\cite{Balram20b} which shows that the 6/17 FQHE state is quite delicate. For this system of $N=12$ electrons, we find that the charge gap (energy required to create a far-separated fundamental quasiparticle-quasihole pair) is less than the neutral gap which indicates the presence of strong finite-size effects. We anticipate that the charge gap would be at least as large as the neutral gap in the thermodynamic limit.  

\section{Low-energy effective theory of the $3\bar{2}1^{3}$ edge}
\label{sec:eff_edge}
In this section, we consider the topological properties of the $3\bar{2}1^{3}$ state that are derived from the low-energy effective theory of its edge. The $3\bar{2}1^{3}$ is the $n=3$ member of the $n\bar{2}1^{3}$ family of states and the edge theory of this sequence of states was worked out in detail in Ref.~\cite{Balram21}. The topological properties of the $3\bar{2}1^{3}$ state are encoded in the $K$ matrix~\cite{Wen91b,Wen92b,Wen95,Moore98}, charge vector $\vec{t}$, and the spin vector $\vec{\mathfrak{s}}$~\cite{Wen92} which are given by~\cite{Balram21}
\begin{equation}
K  =
\begin{pmatrix} 
      2 & 1 & 0 & -1 \\
      1 & 2 & 0 & -1 \\
      0 & 0 &-2 & 1 \\
     -1 &  -1 & 1 & 3\\
   \end{pmatrix},~
   \vec{t}= \begin{pmatrix} 
      0\\
      0\\
      0\\
      1 
   \end{pmatrix}~{\rm and}~
   \vec{\mathfrak{s}}= \begin{pmatrix} 
      -2\\
      -2\\
      1\\
      5/2
   \end{pmatrix}.
\label{eq:Kmatrix_parton_3bar2111}   
\end{equation}
The filling factor and shift can be obtained from the $K$ matrix as~\cite{Wen95} $\nu =  \vec{t}^{\rm T}\cdot K^{-1} \cdot \vec{t} = 6/17$ and $\mathcal{S}=(2/\nu) \left( \vec{t}^{\rm T}\cdot K^{-1} \cdot \vec{\mathfrak{s}} \right)=4$. These values are consistent with that ascertained from the microscopic wave function given in Eq.~(\ref{eq: parton_6_17_3bar2111}). The ground-state degeneracy of the $3\bar{2}1^{3}$ state on a manifold with genus $g$ is~\cite{Wen95} $\mathcal{D} = |{\rm Det}(K)|^{g} =17^{g}$.  The $K$ matrix of Eq.~(\ref{eq:Kmatrix_parton_3bar2111}) has one negative and three positive eigenvalues which implies that the $3\bar{2}1^{3}$ state has a chiral central charge of $2$. 

Following the derivation outlined in Ref.~\cite{Balram21}, one can show that the $K$ matrix, charge and the spin vectors of the CFized $6/5$ state described by the wave function given in Eq. ~(\ref{eq: CFized_6_5}) are as follows:
\begin{equation}
K  =
\begin{pmatrix} 
      3 & 2 \\
      2 & 7\\
   \end{pmatrix},~
   \vec{t}= \begin{pmatrix} 
      1\\
      1 
   \end{pmatrix}~{\rm and}~
   \vec{\mathfrak{s}}= \begin{pmatrix} 
      3/2\\
      9/2
   \end{pmatrix}.
\label{eq: Kmatrix_parton_CFized_6_5}   
\end{equation}
One can readily check that the filling factor and shift obtained from the above $K$-matrix are $\nu = 6/17$ and $\mathcal{S}=4$ that are identical to those determined from the wave function given in Eq. ~(\ref{eq: CFized_6_5}). The ground-state degeneracy of the CFized 6/5 state on the torus is $\mathcal{D} = 17$.  The $K$ matrix of Eq.~(\ref{eq: Kmatrix_parton_CFized_6_5}) has two positive eigenvalues and thus the CFized $6/5$ state has a chiral central charge of $2$.  Remarkably,  these topological properties of the CFized $6/5$ state are identical to those of the $3\bar{2}1^{3}$, indicating that they both likely lie in the same phase.  Thus, it should be possible to view the $3\bar{2}1$ state, which occurs at filling factor $\nu=6/5$, as a $1+1/5$ Laughlin state. We leave out a detailed exploration of this intriguing connection for future work. 

We mention here that the Haldane-Halperin hierarchy state~\cite{Haldane83, Halperin84} at 6/17 that is obtained by condensing Laughlin quasiparticles at 1/3 is also described by the $K$-matrix given in Eq.~(\ref{eq: Kmatrix_parton_CFized_6_5}) and thus likely lies in the same topological class as the $3\bar{2}1^{3}$ state. However, the $6/17$ wave function produced by the hierarchical construction is not readily amenable to a numerical evaluation. Moreover, the 4/11 state, which precedes 6/17 in the hierarchical construction along the sequence $\nu=2n/(6n-1)$ does not lie in the same topological phase as the Coulomb ground state at 4/11~\cite{Balram21}. 

\section{Discussion}
\label{sec: discussions}
In this concluding section, we discuss many experimentally testable properties of the $3\bar{2}1^{3}$ ansatz that can help establish its underlying topological order.  A single particle in the factor $\Phi_{3}$ and $\Phi_{1}$ results in a quasiparticle that carries a charge $-2e/17$ and $-6e/17$ respectively. On the other hand, a single particle in the factor $\Phi_{\bar{2}}$ results in a quasiparticle (obtained by complex conjugating a single hole at $\Phi_{2}$) that carries a charge of $-3e/17$. The smallest charged quasiparticle, generated by creating a hole in the factor of $\Phi_{3}$ and a particle in the factor of $\Phi_{\bar{2}}$, carries a charge of $+2e/17-3e/17=-e/17$. All the excitations of the $3\bar{2}1^{3}$ state are Abelian anyons.

The Hall viscosity $\eta_{H}$ of the $3\bar{2}1^{3}$ state, which is proportional to its shift on the sphere, is quantized~\cite{Read09} as $\eta_{H}=(6/17)~\hbar/(2\pi\ell^{2})$. \emph{Assuming} that all the edge states equilibrate, the thermal Hall conductance $\kappa_{xy}$ at a temperature $T$ that is lower than the gap of the FQHE state is completely determined by the chiral central charge of the FQHE phase~\cite{Kane97}. Therefore, for the $3\bar{2}1^{3}$ state $\kappa_{xy}$ is also expected to be quantized as $\kappa_{xy}=2[\pi^2 k^{2}_{B}/(3h)T]$. Recently, thermal Hall measurements have been carried out at several filling factors in the LLL~\cite{Banerjee17, Srivastav19}. 

The $3\bar{2}1^{3}$ state is the $n=3$ member of the $n\bar{2}1^{3}$ family of states occurring at fillings $\nu=2n/(5n+2)$ with shift $\mathcal{S}=n+1$ and chiral central charge $n-1$. The $n=1$ member is identical to the 2/7 Jain state. The $n=2$ member produces a novel $\mathbb{Z}_{2}$-ordered state at filling factor $1/3$~\cite{Balram20} which although competitive with the 1/3 Laughlin state in the SLL, is unlikely to materialize in the LLL~\cite{Balram20, Faugno20b}. The $4\bar{2}1^{3}$ state provides a candidate wave function at $\nu=4/11$ and was studied in detail in Ref.~\cite{Balram21}. The $5\bar{2}1^{3}$ state occurs at $\nu=10/27$ where as yet no signs of FQHE have been reported.

Finally, we mention the possibility of two-component states at 6/17. In the LLL, we anticipate that the singlet and the partially polarized $6/17$ states likely arise from the corresponding CF states at $\nu^{*}=6/5$~\cite{Balram16c}.  The $3\bar{2}1^{3}$ state also allows for the formation of unpolarized states building on the singlet state at $\nu=2$ and the partially-polarized state at $\nu=3$. It is conceivable that these states could be relevant for certain interactions.

\textit{Acknowledgements} - We acknowledge useful discussions with Maissam Barkeshli. We thank funding support from the Science and Engineering Research Board (SERB) of the Department of Science and Technology (DST) via the Startup Grant No. SRG/2020/000154 (A. C. B) and the Polish NCN Grant No. 2014/14/A/ST3/00654 (A. W.). Numerical calculations reported in this work were carried out on the Nandadevi supercomputer, which is maintained and supported by the Institute of Mathematical Science's High-Performance Computing Center, and the Wroc\l{}aw Centre for Networking and Supercomputing and Academic Computer Centre CYFRONET, both parts of PL-Grid Infrastructure.

\bibliography{../../Latex-Revtex-etc./biblio_fqhe}
\end{document}